# Refractive index and Snell's Law for Dipole-Exchange Spin-Waves in a Confined Planar Structure

Dae-Eun Jeong, Dong-Soo Han, Sangkook Choi, and Sang-Koog Kim*

Research Center for Spin Dynamics & Spin-Wave Devices and Nanospinics Laboratory, Department of Materials Science and Engineering, Seoul National University, Seoul 151-744, Republic of Korea

We derived analytical forms of refractive index and Snell's law for dipole-exchange spin waves of reflection and refraction at a magnetically heterogeneous interface in a geometrically confined planer structure from a microscopic scattering approach. A novel behavior, the optical total reflection, was demonstrated for spin waves with a specific interface between Yttrium iron garnet ($Y_3Fe_5O_{12}$) and Permalloy ($Ni_{80}Fe_{20}$) by analytical and micromagnetic numerical calculations.

Dipole-exchange spin waves (DESWs) in magnetically ordered materials of restricted geometries including one-dimensional (1D) multilayer, 2D patterned rectangular-, circular-, and stripe-shaped magnetic elements have attracted increasing interest in the areas of nanomagnetism and magnetization (**M**) dynamics [1]. Owing to advances in the fundamental understanding of a variety of excited DESW modes in micron or nano-sized magnetic thin films [2], a new class of logic devices [3] along with other integrated electronic circuits [4] based on DESWs have been proposed. An understanding of the macroscopic wave properties of DESWs traveling in magnetic waveguides of lateral confinements is a prerequisite for making such DESW's devices technologically applicable.

In principle, macroscopic wave behaviors in optics are the results of microscopic interactions between optical waves and charged particles in optical media, the so-called scattering phenomenon. Accordingly, Snell's law in optics can be described in terms of the refractive indices of different optical media. Analogously, the macroscopic wave behaviors of DESWs can also be expounded simply with reference to the DESW's refractive index of a given magnetic medium, and the DESW's refractive index can be expressed in terms of magnetic material parameters. Consequently, Snell's law described in terms of the DESWs refractive indices, gives rise to analytic constraints of the wave behaviors at the interface between different magnetic media, and moreover, would provide much quantitative information on their wave behaviors of reflection and refraction in thin-film magnetic media[5,6].

In this Letter, Snell's law of reflection and refraction for DESWs in laterally restricted geometry of magnetic thin films, along with an explicit form of the refractive index for DESWs in a different magnetic material were analytically derived from a microscopic approach using both the dispersion relations of two distinct magnetic media and a specific boundary condition at their interface. The analytical derivations were also confirmed by micromagnetic numerical calculations conducted on, not only the macroscopic refraction and reflection behaviors of DESWs, but also their total reflection at a specific interface between Permalloy (Py: $Ni_{80}Fe_{20}$) and Yttrium iron garnet (YIG : $Y_3Fe_5O_{12}$).

To obtain the analytical forms of the refractive index and Snell's law for DESWs, we begin with the Landau-Lifshitz (LL) equation governing dynamic **M** motion within a continuum approximation, neglecting the phenomenological damping. The LL equation is expressed as $\partial \mathbf{M}/\partial t = -\gamma \left( \mathbf{M} \times \mathbf{H}_{eff} \right)$ where $\mathbf{H}_{eff}$ is the effective field and $\gamma$ the gyromagnetic ratio [7]. Since the DESW is a collective excitation of constituent **M**s in restricted geometry, both of the exchange and dipolar interactions significantly contribute to $\mathbf{H}_{eff}$. To describe collective **M** motions in wavelike forms, we separated the static and dynamic terms of **M**s, such that $\mathbf{M}(\mathbf{x},t) = \mathbf{M_0} + \mathbf{m}(\mathbf{x},t) = M_0 \hat{y} + \mathbf{m}(\mathbf{x})e^{-i\omega t}$, where $\hat{y}$ is the unit vector along the $y$ axis, $\mathbf{M_0}$ is the static **M** aligned in the $+\hat{\mathbf{y}}$ direction (see Fig. 1). In analogy, the magnetic field was also separated into the static and dynamic terms, such that $\mathbf{H}^{int}(\mathbf{x},t) = \mathbf{H}_0^{int} + \mathbf{h}(\mathbf{x})e^{-i\omega t}$. In a plane-wave approach with a time harmonics, the dynamic motion of DESWs with a given wavevector **k** and angular frequency $\omega$ in a soft ferromagnet can be described as [8, 9]

$$-i\omega \mathbf{m}(\mathbf{k},\omega) = \gamma \mathbf{M_0} \times \left\{ \mathbf{h}(\mathbf{k},\omega) - \left( \frac{A_{ex} a l_{ex}^5}{8(S\mu_B)^2} k_i k_j + \frac{\mathbf{M_0} \cdot \mathbf{H}_0^{int}}{M_S^2} \right) \mathbf{m}(\mathbf{k},\omega) \right\}, \quad (1)$$

where $A_{ex}$ is the exchange stiffness, $a$ is the lattice parameter, $l_{ex}$ is the exchange length, $S$ is the magnitude of spin, $\mu_B$ is the Bohr magneton, and $M_s$ is the saturation **M**. $\mathbf{H}_0^{int}$ is the static internal field inside the model sample, in this study is given along the y axis. The subscripts in k indicate the x, y, z coordinates [10]. We assume the magnetocrystalline anisotropy to be zero for such a soft ferromagnet.

To investigate the macroscopic wave behaviors of DESWs, next it is necessary to calculate the value of $\mathbf{H}_0^{int}$ intrinsically given to the model geometry of lateral confinements. To do this, we used tensorial Green's functions that allow for an analytical form for $\mathbf{H}_0^{int}$ through integration over the volume of a finite size. For a rectangular-shaped thin film, $\mathbf{H}_0^{int} \cong \mathbf{H}^{appl} - N_{yy}\mathbf{M}_S$ with an applied field $\mathbf{H}^{appl}$ and the demagnetization factor $N_{yy}$. $N_{yy}\mathbf{M}_S = \int \mathbf{G}(\mathbf{r},\mathbf{r}') \cdot \mathbf{M}(\mathbf{r}',t) d\mathbf{r}'$ is calculated using $\mathbf{G}(\mathbf{r},\mathbf{r}') = -\frac{1}{2\pi}\frac{\partial^2}{\partial y^2} R(|\mathbf{r}-\mathbf{r}'|/L)\hat{y}$ with $R(X) = \sinh^{-1}(1/X) + X - \sqrt{1+X^2}$ [11] and is thus determined by the specific geometry and the magnetization state of a given system, in this study e.g. by the rectangular thin film being saturated with $M_0$ in the $+y$ direction, as noted by the black arrow in Fig. 1.

In addition to the dynamic equation of **M** motion, a relationship between $\mathbf{m}(\mathbf{k},\omega)$ and $\mathbf{h}(\mathbf{k},\omega)$ is required to obtain the $\omega$ - **k** relation of DESWs. For the case of $\omega/k \ll c$ with the light velocity $c$, the magnetic field is quasi-static [12], and under this approximation we have a Maxwell equation $\mathbf{k}\cdot\mathbf{h}(\mathbf{k},\omega) = -\mathbf{k}\cdot\mathbf{m}(\mathbf{k},\omega)$. Inserting this relation into Eq. (1), we have

$$\left(k^2\right)^2 + \frac{8S^2\mu_B^2}{A_{ex}al_{ex}^5}\left(\sin^2\theta - 2N_{yy} + \frac{2H_y^{appl}}{M_S}\right)k^2$$
$$+ \frac{64S^4\mu_B^4}{A_{ex}^2a^2l_{ex}^{10}}\left(\frac{(H_y^{appl})^2}{M_S^2} + \left(\sin^2\theta - 2N_{yy}\right)\frac{H_y^{appl}}{M_S} + N_{yy}^2 - \sin^2\theta N_{yy} - \frac{\omega^2}{\gamma^2 M_S^2}\right) = 0 \quad , \quad (2)$$

where $\theta$ is an arbitrary angle of **k** with respect to the $+\hat{\mathbf{y}}$ direction. For a decay-free mode, the nontrivial solution of Eq. (2) is a function of both variables $\theta$ and $\omega$, and is given simply as:

$$k = \frac{2S\mu_B}{\left(Aal_{ex}^5\right)^{1/2}}\left[\left(2N_{yy} - \sin^2\theta - \frac{2H_y^{appl}}{M_S}\right) + \left(\sin^4\theta + \frac{4\omega^2}{\gamma^2 M_S^2}\right)^{1/2}\right]^{1/2} . \quad (3)$$

When the decay-free DESW mode expressed by Eq. (3) propagates from one medium to the other through their interface, DESW's reflection at and refraction through an interface between different magnetic media can occur. For the explicit derivation of Snell's law for DESWs, we consider an effective pinning boundary condition for the inhomogeneity of the dynamic **M**s at the interface. This condition requires the continuity of the tangential *k* components of incident, reflected, and refracted DESWs at the specific interface. Note that the electrodynamic boundary conditions typically employed in wave optics are no longer sufficient when considering both the exchange and dipolar interactions in confined magnetic elements of micrometer or smaller in size. An additional boundary condition should be considered and it is thus obtained through the integration of the equation of dynamic **M** motion at the interface. The torque by the exchange coupling, the interface magnetocrystalline anisotropy [13], and the effective pinning by the inhomogeneity of the dynamic demagnetization field at the interface [14] together make a specific boundary condition, such that $\partial \mathbf{m}/\partial y - d_1 \mathbf{m}|_{y=0} = \partial \mathbf{m}/\partial y - d_2 \mathbf{m}|_{y=0}$, where $d_{1,2}$ is the effective pinning parameter, and is a function of $d_S = K_S/A_{ex}$ and $d_D$, where $K_S$ is the interface magnetocrystalline anisotropy constant and $d_D$ is the dipole pinning term, given as $d_D = 2\pi/\eta \left[1 + 2\ln(1/\eta)\right]$ with the ratio of thickness to width, $\eta$. For a soft magnet $d_S$ is negligible. Thus, the dynamic term of $\mathbf{M}(\mathbf{x},t)$, i.e., $\mathbf{m}(\mathbf{x})e^{-i\omega t}$ at the interface in a thin film magnet of its finite lateral confinement, being magnetized in the film plane can be described only by $d_D$, which is of a pure dipolar nature. In addition, there is another torque term acting on **M**s at the interface. Under an assumption that the intercoupling at the interface[15] is relatively weak compared to that inside each medium, the boundary condition yields the continuity of the tangential **k** components of the incident, reflected and refracted DESWs, such that $k_\parallel^0 = k_\parallel' = k_\parallel''$. In analogy to wave optics, the dimensionless refractive index of a magnetic medium is finally obtained as

$$n_M(\theta,\omega) = \frac{2S_p \mu_B c}{\omega (A a l_{ex}^5)^{1/2}} \left[\left(2N_{yy} - \sin^2\theta - \frac{2H_y^{appl}}{M_S}\right) + \left(\sin^4\theta + \frac{4\omega^2}{\gamma^2 M_S^2}\right)^{1/2}\right]^{1/2}. \qquad (4)$$

This analytical form of $n_M$ for an arbitrary soft magnetic medium characterizes the macroscopic wave behavior of DESWs propagating in a given medium. In elastic scattering process, a selected value of $\omega$ remains unchanged after interacting with the interface, so that Snell's law for DESWs can be expressed as $n_{M1} \sin\theta_{M1} = n_{M2} \sin\theta_{M2}$ for magnetic media 1 and 2, through the $\omega$ versus $\mathbf{k}$ and the $k_\parallel^0 = k_\parallel' = k_\parallel''$ relations. On the basis of Snell's law, all the aspects of macroscopic reflection and refraction phenomena for DESWs can be described/interpreted simply in terms of $n_M$.

Figure 2(a) shows the numerical calculations, versus both $\theta$ and $f_{SW}$, of the analytical equation of $n_M$ for YIG and Py magnets, with an application of $H_y^{appl} = 1.0$ T. The numerical results offer a useful information on the macroscopic wave properties of Py and YIG media, such that $n_{Py} > n_{YIG}$. Py is denser than YIG with reference to SW propagations. In other word, as the DESWs propagate from a magnetically rare medium to a dense one, e.g., from YIG to PY, the angle-of-incidence is greater than the angle-of-refraction, In Fig. 2(b), the relation of $\theta_{YIG}^0$ and $\theta_{Py}''$ was plotted according to Snell's law of DESWs, $n_{YIG} \sin\theta_{YIG}^0 = n_{Py} \sin\theta_{Py}''$ for a specific frequency, here $f_{SW} = 51$ GHz: $\theta_{YIG}^0 > \theta_{Py}''$ is maintained since $n_{Py} > n_{YIG}$. For the other case where SWs are incident from Py to YIG, the angle-of-refraction $\theta_{YIG}''$ is greater than the angle-of-incidence $\theta_{Py}^0$. In this case, an internal total reflection can happen, as in wave optics, above a critical incidence angle given as $\theta_{Py,cri}^0 = \sin^{-1}\left(n_{YIG}/n_{Py}\right)$, 15.16 degree.

Next, to confirm the validity of the macroscopic parameter of $n_M$ and the DESW's Snell's law derived from a microscopic scattering approach, we conducted micromagnetic numerical simulations [16] on a model geometry composed of two different magnetic media with their interface, as shown in Fig. 3. The two media, e.g., YIG (y < 0 nm) [17] and Py (y > 0 nm) [18] were saturated in the +y direction with an application of the external magnetic field of 1 T. For the generation of monochromatic ($f_{SW} = 51$ GHz) DESWs with flat wavefront normal to the 45° incidence angle, the oscillating magnetic field, $\mathbf{H}_{osc} = A_H (1 - \cos 2\pi v_H t)\hat{\mathbf{x}}$ with $A_H$ =250 Oe and $v_H$ = 51 GHz, was applied only to a local area indicated by the dash-dot boxed region [5] as shown in Fig. 1. Figure 3(a) shows the perspective-view snapshot images of the temporal evolution of the spatial distribution of the out-of-plane components of the local **M**s normalized by the saturation **M**, i.e., $M_z / M_S$, in the central area including the Py/YIG interface (marked by the gray-colored box in Fig. 1). When the DESWs traveling initially in the YIG medium encounter the YIG / Py interface, some of them are reflected, thus forming interference patterns by the superposition of the incident DESWs and reflected ones in the YIG medium. Some of the initially traveling DESWs continue to propagate with an refraction angle $\theta''_{Py}$ in the other Py medium through the interface, as a well-known optical wave's refraction.

For further quantitative analysis, we made the fast-Fourier-transform (FFT) of the temporal evolution of spatial $M_z / M_S$ distributions. Figure 3(b) shows the results where three strong peaks appear at $k_x = 0.0310$ nm$^{-1}$, as indicated by the horizontal line. These three peaks, positioned at $k_y = -0.0301$, $0.0310$, and $0.1396$ nm$^{-1}$ are identified to be (as) the reflected (blue), incident (red), and refracted (green) rays of DESWs, respectively [19], from the frequency contours of the allowed DESWs in the Py (purple dotted line) and YIG (orange dotted line) media, obtained from the analytical Eq. (3) for a specific frequency, $f_{SW} = 51$ GHz. This result reveals the fact that the incident, reflected, and refracted DESWs have an equal value of $k_\parallel$, here $k_x = 0.0310$ nm$^{-1}$. This agreement between the analytic and micromagnetic numerical calculations verifies that DESWs obey Snell's laws of reflection and refraction as in wave optics..

Moreover, we demonstrated a total reflection behavior for DESWs by micromagnetic simulations on the same model geometry as in Fig. 1, but the medium 1 and 2 correspond to YIG and Py, respectively, and with a different value of $v_H = 60$ GHz. Figure 4(a) shows snapshot images of the resultant spatial distribution of the $M_z / M_S$ components taken at the indicated several times. As DESWs were incident on the interface, all of them were reflected, thus forming certain interference patterns in the Py medium, whereas there is no refracted ray in the YIG medium. The FFT power contour on the x −y plane of the simulation results, shown in Fig. 4(b), reveals that there is only a strong peak at $(k_x, k_y)= (0.105$ nm$^{-1}$, $0.105$ nm$^{-1})$ and a very blurred peak at $(0.105$ nm$^{-1}$, $-0.105$ nm$^{-1})$. It is also confirmed that the orange-(Py) and purple-color dotted curves (YIG) obtained from analytical calculations allow for identifying the incident (red) and reflected (blue) rays of DESWs [19], in accordance with the reflection constraint of $k_\parallel^0 = k_\parallel'$, In contrast, there were no refracted DESWs in the YIG medium according to the refraction constraint of $k_\parallel^0 = k_\parallel''$. These results together indicate that optical wave's total reflection can possibly take place for DESWs in some specific conditions.

In conclusion, we explicitly derived refractive index and Snell's law of reflection and refraction for dipole-exchange spin-waves in a geometrically confined planer structure composed of different magnetic media. The macroscopic wave properties of reflection and refraction for DESWs are analytically described and interpreted in terms of a dimensionless parameter, the refractive index of a given magnetic medium and the corresponding Snell's law. The analytical calculations are also verified by micromagnetic numerical calculations of a total refection as well as the reflection and refractive behaviors of DESWs at a specific interface between Py and YIG, for example.

We thank S. O. Demokritov for his valuable comments and careful reading of this manuscript. This work was supported by Creative Research Initiatives (ReC-SDSW) of MEST/KOSEF.

[15] F. Hoffmann, Phys. Status Solidi **41**, 807 (1970).

[16] In the present micromagnetic simulations, we used the OOMMF code (version 1.2a4). See http://math.nist.gov/oommf., The cell size and the Gilbert damping parameter used are $5\times5\times5$ nm$^3$ and 0.01, respectively.

[17] For YIG, we used $A_{ex} = 3\times10^{-12}$ J/m, $M_S = 1.27\times10^4$ A/m, and $K = 0$ J/m$^3$.

[18] For Py, $A_{ex} = 1.3\times10^{-11}$ J/m, $M_S = 8.6\times10^5$ A/m, and $K = 0$ J/m$^3$.

[19] The frequency contours were obtained for the case of the lowest DESW mode of the quantization number along the film thickness, i.e., $n = 0$. For details, see B. A. Kalinikos, in *Dipole-Exchange Spin-Wave Spectrum of Magnetic Films*, edited by M. G. Cottam, Linear and Nonlinear Spin Waves in Magnetic Films and Superlattices, (World Scientific, Singapore, 1994);

**Figure captions**

Fig.1. (online color) Laterally confined thin–film geometry of the model system composed of two different magnets of 5.0 nm thickness, both media being saturated with $M_S$ in the $+y$ direction as indicated by the black-colored arrow. The dash-dot box is an area for local excitation of monochromatic DESW modes with flat wave-front normal to the incidence angle $\theta^0$ from medium 1. Gray-colored square area is a selected area for examining DESW's reflection and refraction. $\hat{k}^0$, $\hat{k}'$ and $\hat{k}''$ indicate the unit wavevectors of the incident, reflected and refracted DESWs, respectively.

Fig. 2. (online color) (a) Color contour plots of the refractive indices of YIG and Py on the $\theta$ - $f_{SW}$ plane, obtained using the analytical Eq. (4). (b) Calculation of the angle of refraction $\theta''_{Py}$ versus the angle of incidence $\theta^0_{YIG}$ for a specific value of $f_{SW} = 51.0$ GHz.

Fig. 3. (online color) (a) Serial snapshot images of the local $M_z/M_S$ distribution taken at the indicated times, obtained by micromagnetic simulations on the 5 nm-thick thin-film model composed of YIG (y < 0 nm) and Py (y > 0 nm), as shown in Fig. 1. (b) FFT power distribution on the $k_x$-$k_y$ plane, obtained from the local $M_z / M_S$ distribution on the $x$-$y$ plane. The purple- and orange-colored dotted curves represent the frequency contours of the DESW modes in the Py and YIG media, respectively, which were obtained from the numerical solution of the analytical Eq.(3).

Fig. 4. (online color) (a) Serial snapshot images of the local $M_z$ / $M_S$ distribution taken at the indicated times, with a different model, Py (y < 0 nm) and YIG (y > 0 nm) media with the same 5nm thickness. (b) FFT power distribution on the $k_x$-$k_y$ plane, along with the frequency contours numerically calculated using the analytical Eq. (3).

Fig. 1

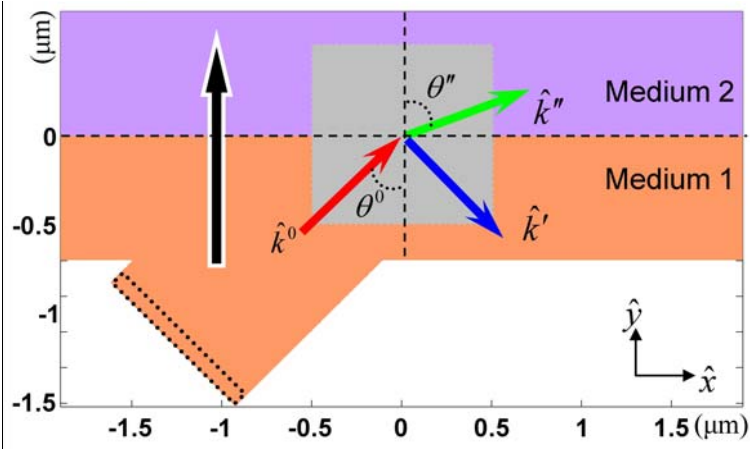

Fig. 2

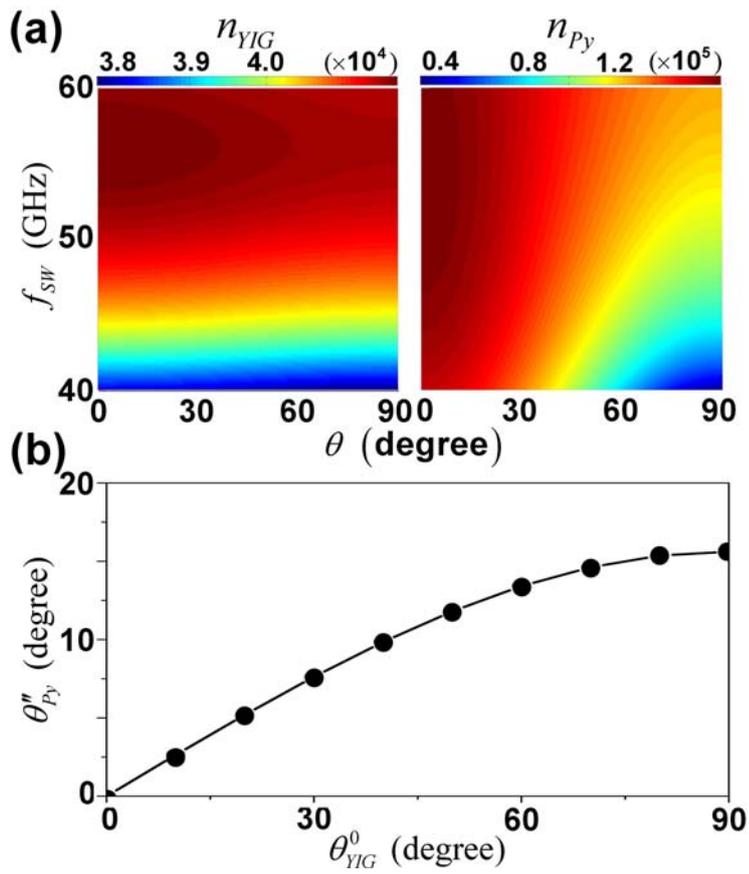

Fig. 3

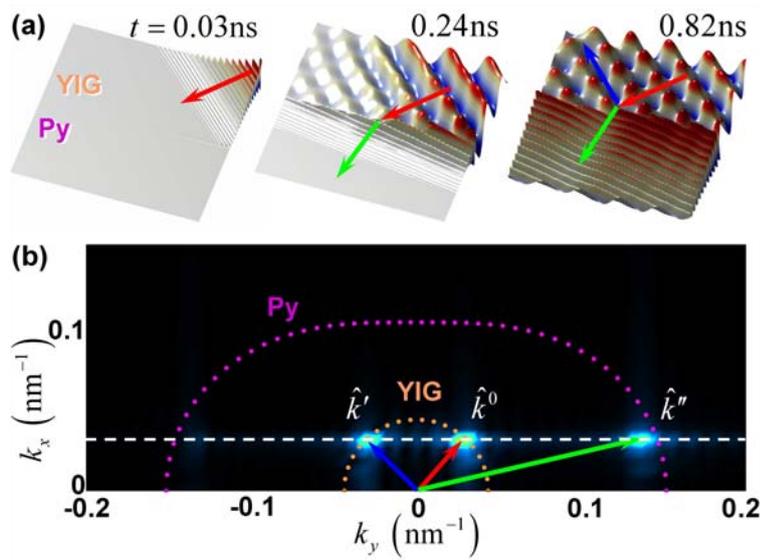

Fig. 4

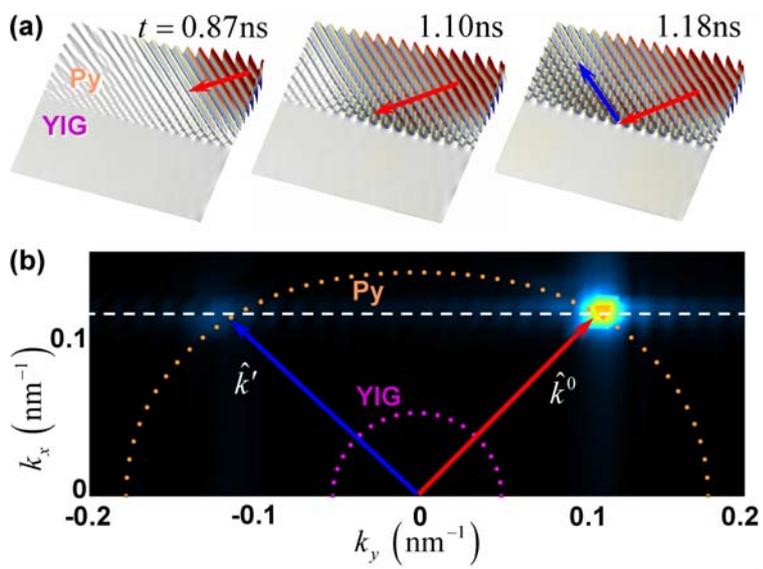